\documentclass[rnote]{aa} % for the research notes
\usepackage{graphicx}
\usepackage{txfonts}
\usepackage[round,authoryear]{natbib}
\begin{document}
   \title{The $\Sigma-D$ Analysis of Recently Detected Radio Planetary Nebulae in the Magellanic Clouds}

   \author{B. Vukoti\' c
          \inst{1}\fnmsep\thanks{Corresponding author}
	  ,
	  D. Uro\v sevi\' c
	  \inst{2, 4}
	  ,
	   M.D. Filipovi\' c
	  \inst{3}
          \and
          J.L. Payne\inst{3}
          }

   \institute{Astronomical Observatory Belgrade,
              Volgina 7, 11160 Belgrade-74, Serbia\\
              \email{bvukotic@aob.bg.ac.yu}
         \and
             University of Belgrade, Faculty of Mathematics, Studentski trg 16, 11000 Belgrade, Serbia\\
             \email{dejanu@matf.bg.ac.yu}
         \and
	     University of Western Sydney, Locked Bag 1797, Penrith South, DC, NSW 1797, Australia\\
	     \email{m.filipovic@uws.edu.au}\\
	     \email{snova4@msn.com}
	 \and {Isaac Newton Institute of Chile, Yugoslavia Branch, Yugoslavia}
             }

   \date{Received February 06, 2009; accepted ???}

\authorrunning{Vukoti\' c et al.}
\titlerunning{The $\Sigma-D$ analysis of Recently Detected Radio PNe in the MCs}

% \abstract{}{}{}{}{} 
% 5 {} token are mandatory
 
  \abstract
   {}
   {To investigate and analyze the radio surface brightness to diameter ($\Sigma-D$) relation  for recently detected, bright radio-continuum planetary nebulae (PNe) in the Magellanic Clouds (MC).}
   {We apply a Monte Carlo analysis in order to account for sensitivity selection effects on measured $\Sigma-D$ relation slopes for bright radio PNe in the MCs.}
   {In the $\Sigma-D$ plane these radio MCs PNe are positioned among the brightest of the nearby Galactic PNe, and are close to the $D^{-2}$ sensitivity line of the MCs radio maps. The fitted Large Magellanic Cloud (LMC) data slope appears to be influenced with survey sensitivity. This suggests the MCs radio PN sample represents just the ``tip of the iceberg'' of the actual luminosity function. Specifically, our results imply that sensitivity selection tends to flatten the slope of the $\Sigma-D$ relation. {Although MCs PNe appear to share the similar evolution properties as Galactic PNe, small number of data points prevented us to further constrain their evolution properties.}}
   {}


   \maketitle
%
%________________________________________________________________

\section{Introduction}
 	
  \subsection{Radio bright Magellanic Cloud planetary nebulae}
\citet{filipovic_etal09} recently reported {15} radio-continuum PNe detected at the $3\sigma$ and above level from various radio Magellanic Clouds (MCs) surveys. Their detections are mainly based on positional coincidences with optically detected Planetary Nebulae (PNe) {and optical spectroscopy \citep{2008SerAJ.176...65P,2008SerAJ.177...53P}. Included are}  Large Magellanic Cloud (LMC) mosaics \citep{2005AJ....129..790D} having sensitivities of $\sim${0.5} mJy~beam$^{-1}$ at both 4.8 and 8.64~GHz with resolutions of 33 and 20\arcsec, respectively. The Small Magellanic Cloud (SMC) mosaics have 0.5 mJy~beam$^{-1}$ sensitivities at 4.8 and 8.64~GHz with resolutions of 30 and 15\arcsec\ {\citep{dickel-iau}}. To compare these sources with Galactic PNe, we use radio surface brightness ($\Sigma$) since this quantity is distance independent:
\begin{equation}
 \Sigma[\mathrm{W} \mathrm{ m^{-2}~Hz^{-1}~sr^{-1} }]=1.505\times10^{-19}\frac{S[\mathrm{Jy}]}{\theta^2[\mathrm{'}]},
\end{equation}
where $S$ is the flux density and $\theta$, the angular diameter of the source. Theory predicts the existence of a linear relation between $\log \Sigma$ and $\log D$, with $D$ representing the diameter of the object \citep[see][and references therein]{2007SerAJ.174...73U,2009A&A...495..537U}.

\begin{figure*}
\centering
 \includegraphics[angle=0,height=0.75\textheight]{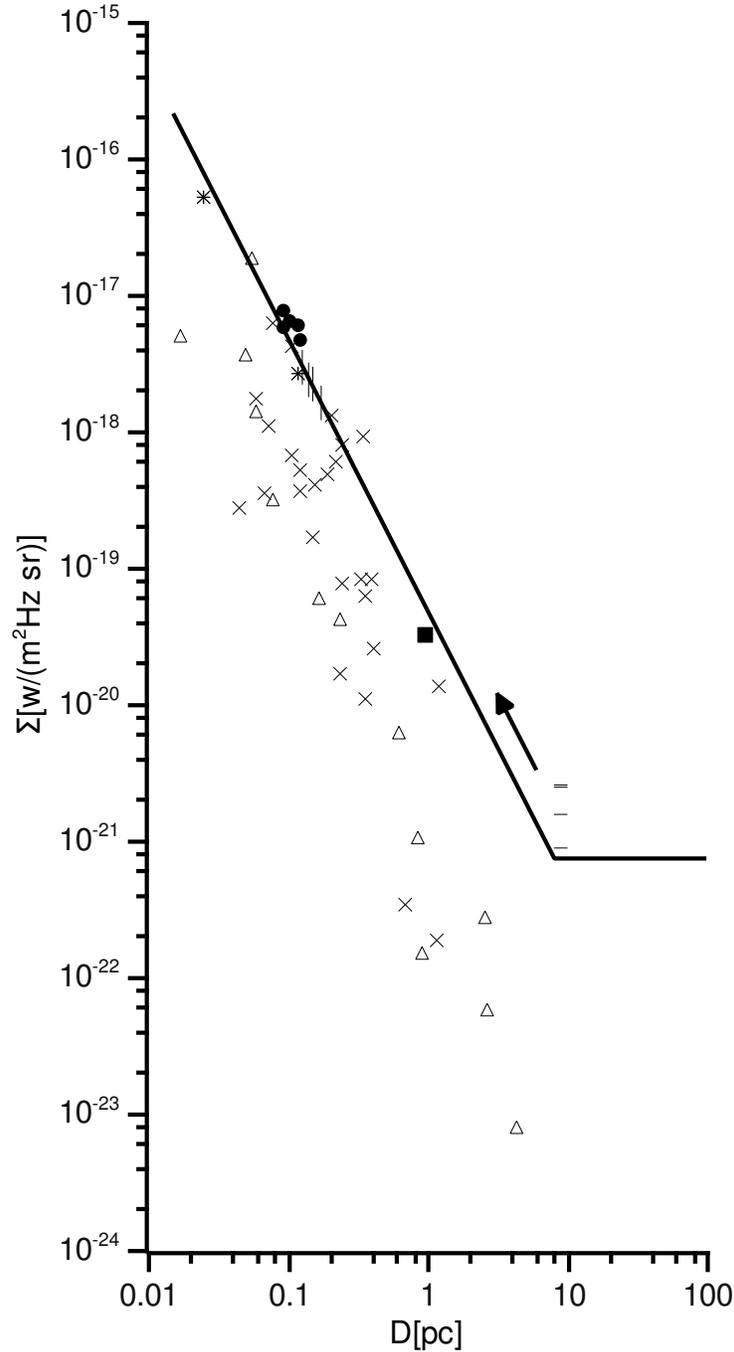}
\caption{A plot of 4.8~GHz PNe from the LMC, SMC and Sagittarius Dwarf Galaxy.
The solid line {represents} the LMC map sensitivity ($3\sigma$) {at 1.5~mJy~beam$^{-1}$}. {The LMC radio PNe with known optical diameters are represented with filled circles (above the sensitivity line, 5 PNe) and vertical dashes (sensitivity line is the upper flux density limit, 5 PNe). One LMC PN observed in ''snap-shot'' mode is represented with filled square.} The SMC radio PNe without known optical diameters are represented by {horizontal} dashes and asterisks are two Sagittarius Dwarf PNe. Open triangles represents Galactic SSV PNe with distances greater than 1~kpc while diagonal crosses are those with distances less than 1~kpc. Note that the arrow represents the direction in which {MCs} radio PNe should move on the graph, if their diameters were known.}
\label{Fig1}
\end{figure*}

All of the detected radio PNe are unresolved, {despite five of them being observed in ``snap-shot'' mode with the resolution of up to 1\arcsec and sensitivity of $\sim0.1~\mathrm{mJy~beam^{-1}}$ \citep[for details see][]{filipovic_etal09}. The} optical diameters are available for some {10} out of {15} \citep[see Table~1 in][]{filipovic_etal09}. The MCs PNe data at 4.8~GHz are plotted in Fig.~\ref{Fig1} with respective LMC map $3\sigma$ sensitivity line at {$1.5~\mathrm{mJy~beam^{-1}}$} (hereafter referred to as sensitivity line). {Five LMC PNe are above the sensitivity line and one is below (``snap-shot'' mode). Five PNe with estimated upper flux density limit of $1.5~\mathrm{mJy~beam^{-1}}$ are plotted on the sensitivity line and the remaining one without optical diameter is not plotted. Four SMC PNe are plotted on the SMC map resolution limit of 30\arcsec.} We use the PNe sample of Galactic PNe with reliable individual distances from \citeauthor*{2008ApJ...689..194S} \citetext{\citeyear{2008ApJ...689..194S}; hereafter \ SSV}, and divide them into two subsamples (nearby PNe with distances smaller than 1~kpc and those with distances larger than 1~kpc). \citet{2009A&A...495..537U}, in an extensive analysis of the empirical $\Sigma-D$ relation for PNe, argued that the SSV sample of nearby PNe is least influenced by selection effects. For comparison, we plot two radio PNe from the Sagittarius dwarf galaxy \citep{2000A&A...363..717D}, also representative of extragalactic radio PNe. An arrow in Fig.~\ref{Fig1} represents the $D^{-2}$ direction in which unresolved objects should move if their diameters were known.

Inspection of Fig.~\ref{Fig1} reveal that the detected radio-continuum MCs PNe are very close to their sensitivity lines and they are as bright as the brightest Galactic PNe. {When compared with Galactic PNe it appears that current data set is only the very peak of the actual MCs PNe luminosity function.} In this paper, {using $\chi^2$ analysis and Monte Carlo generated data samples}, we {investigate the sensitivity related selection effects on}  the $\Sigma-D$ properties of these objects {and compare them} with the properties of Galactic PNe. 

 \subsection{Radio planetary nebulae sample selection effects - a brief history}

As a convenient platform for distance determination, the $\Sigma-D$ relation\footnote{In this paper, we use the form $\Sigma\propto D^{\beta}$, where $\beta$ represents slope.} has a $\sim$50~year long history of exploitation for supernova remnants (SNRs). The study of the relation for these very luminous radio synchrotron sources began with the work of \citet{1960AZh....37..256S}, while the first empirical $\Sigma-D$ relation was presented by \citet{1968AJ.....73...65P}. Decades of analysis resulted in a good understanding of the \mbox{$\Sigma-D$} relation for SNRs and various data sample selection effects that influence that relation.

The previous analysis of  PNe radio data samples \citep[e.g.][and references therein]{2002ApJS..139..199P} were mainly focused on radius vs radio brightness temperature ($R-T_\mathrm{b}$) empirical relations. These were sparse in the consideration of data selection effects. Recent works by \citet{2007SerAJ.174...73U,2009A&A...495..537U} use a SNR formalism in the analysis of PNe data along with a discussion of these selection effects. Although they note that all PN samples suffer from selection effects caused by limitations in survey sensitivity and resolution, the nearby SSV sample, having a $\Sigma-D$ slope of ~$\beta=-2.61\pm0.21$, is least influenced by these effects.

\section{Analysis}

 \subsection{Planetary nebulae data sample sets}
 
Samples for Monte Carlo simulations are made using sources from Table~1 in \citet{filipovic_etal09} that have a known optical diameter (LMC sources only) {and are above the map sensitivity line for a given frequency}. This includes {5 PNe at both frequencies (8.64~ and 4.8~GHz) though not necessarily the same ones}. Using adopted distances to the LMC of 50~kpc and to the SMC of 60~kpc \citep{2004NewAR..48..659A,2005MNRAS.357..304H}, the smallest source{s} in the selected sample{s} have a {mean geometrical} diameter of {0.09}~pc at both frequencies. The largest source {has a mean geometrical diameter of 0.12~pc at 4.8~GHz and 0.15~pc at 8.64~GHz}. Surface brightness ranges from ${2.67} \times 10^{-18}$ to $8.65 \times 10^{-18}$ Wm$^{-2}$~Hz$^{-1}$~sr$^{-1}$ at 8.64~GHz and from ${4.70} \times 10^{-18}$ to $7.90 \times 10^{-18}$  Wm$^{-2}$~Hz$^{-1}$~sr$^{-1}$ at 4.8~GHz. For the LMC map{s}, the 4.8~GHz sensitivity line is at $\Sigma= {7.46} \times 10^{-22}$ Wm$^{-2}$~Hz$^{-1}$~sr$^{-1}$ with a $D^{-2}$ break at $D=8.00$~pc and a 8.64~GHz sensitivity line at $\Sigma={2.03} \times 10^{-21}$ and $D=4.85$~pc, respectively.

The relative flux density errors at both frequencies of the LMC maps are $<$10\% { \citep{filipovic_etal09} and they are approximately the size of the symbols shown in Fig.~\ref{Fig1}}. If we assume that the optical diameters have no significant errors and that the absolute flux density error equals the flux density standard deviation $\sigma_\mathrm{S}$, according to error propagation theory it follows from Eq.~1. that the standard deviation for $\log \Sigma$ ($\sigma_\mathrm{log\Sigma}$) is $0.434~\Delta S/S$. This gives \mbox{$\sigma_\mathrm{log\Sigma}\approx 0.05$.} 

The resulting parameters of the $\log \Sigma=a+\beta* \log D$ fits are given in Table~\ref{fits} for each frequency. The parameters $a$ and $\beta$, their standard uncertainties $\Delta a$ and $\Delta \beta$, the linear correlation coefficient $r$, the fit quality $r^2$, the probability $Q$ of obtaining larger WSSR (weighted sum of square residuals) and a ratio of WSSR to the number of degrees of freedom (ndof) are given in this table. A value of {$0.1\gtrsim Q \gtrsim0.001$} is expected for a statistically acceptable fit {(the case when errors of the dependent variable have a non-Gaussian distribution)}, if the data is were well approximated by the model. This should apply for fits presented in this paper since we have transformed flux errors to $\log \Sigma$ errors. When introduced in the  least squares fitting procedure, our values of $Q$ implies a statistically acceptable fit. 

{Although the fits are statistically believable the resulting $\beta$ at 4.8 GHz substantially differs from $\beta$ at 8.64 GHz, likely in debt to a small number of data points.} {Sampling} only the peak of the luminosity function {and} with the majority of the sample likely hidden below the sensitivity line, the resulting values for $\beta$ should not be taken seriously. {Comparing} the LMC radio PN sample and the sample of a nearby Galactic PNe, we made an attempt to access a more meaningful $\Sigma-D$ slope using Monte Carlo simulations as described below.

\begin{table*}
\centering
\caption{Fit parameters of the LMC data sample.} \label{fits}
\begin{tabular}{lcccccccc}
\noalign{\smallskip}
\hline \hline
\noalign{\smallskip}
 Frequency&$a$&$\Delta a$&$\beta$&$\Delta \beta$&$r$&$r^2$[\%]&$Q$&$\mathrm{\frac{WSSR}{ndof}}$\\
\noalign{\smallskip}
\hline \noalign{\smallskip}
8.64  GHz&--19.7&0.3&--2.6&0.4&--0.94&88.36&$5.02\times 10^{-2}$&2.60\\
\noalign{\smallskip}
4.80   GHz&--18.1&0.5&--0.9&0.5&--0.64&40.96&$9.96\times 10^{-2}$&2.08\\
\hline \noalign{\smallskip}
 \end{tabular}
 \end{table*}

 \subsection{Monte Carlo simulations}
 
We performed Monte Carlo simulations similar to those described by \citet{2005A&A...435..437U}. {The results of their simulations  (with 21 data points above the sensitivity line) showed that slopes steeper than $-2$ are under influence of sensitivity related selection effects}.  First, we determined the empirical $\log\Sigma$ standard deviation from the best fit line, assuming $\log D$ as the independent variable. {We then selected an interval in $\log D$ five times as long as that of the real data. This interval is then} sprinkled with random points of the same $\log D$ density as that of the real data.

The simulated points, that lie on the log D axis, are then projected onto a series of lines at different slopes (in steps of 0.1 from --3.5 to --1.5). Each of these lines passes through the extreme upper left hand end of the best fit line to the real data. We also added Gaussian noise
in $\log\Sigma$, which is related to the scatter of the real data by a parameter called ``scatter''. A scatter of 1 corresponds to the same standard deviation as that of the real data.

An appropriate sensitivity cutoff is applied to the simulated data points, selecting points above the sensitivity line. This is done 100 times for each simulated slope and a least squares best fit line is generated for artificial samples that have five or more selected data points. The number of such samples is given in the last column of Tables~\ref{48ghz} and~\ref{864ghz}.
 
\begin{figure*}
\centering
 \includegraphics[angle=0,height=0.75\textheight]{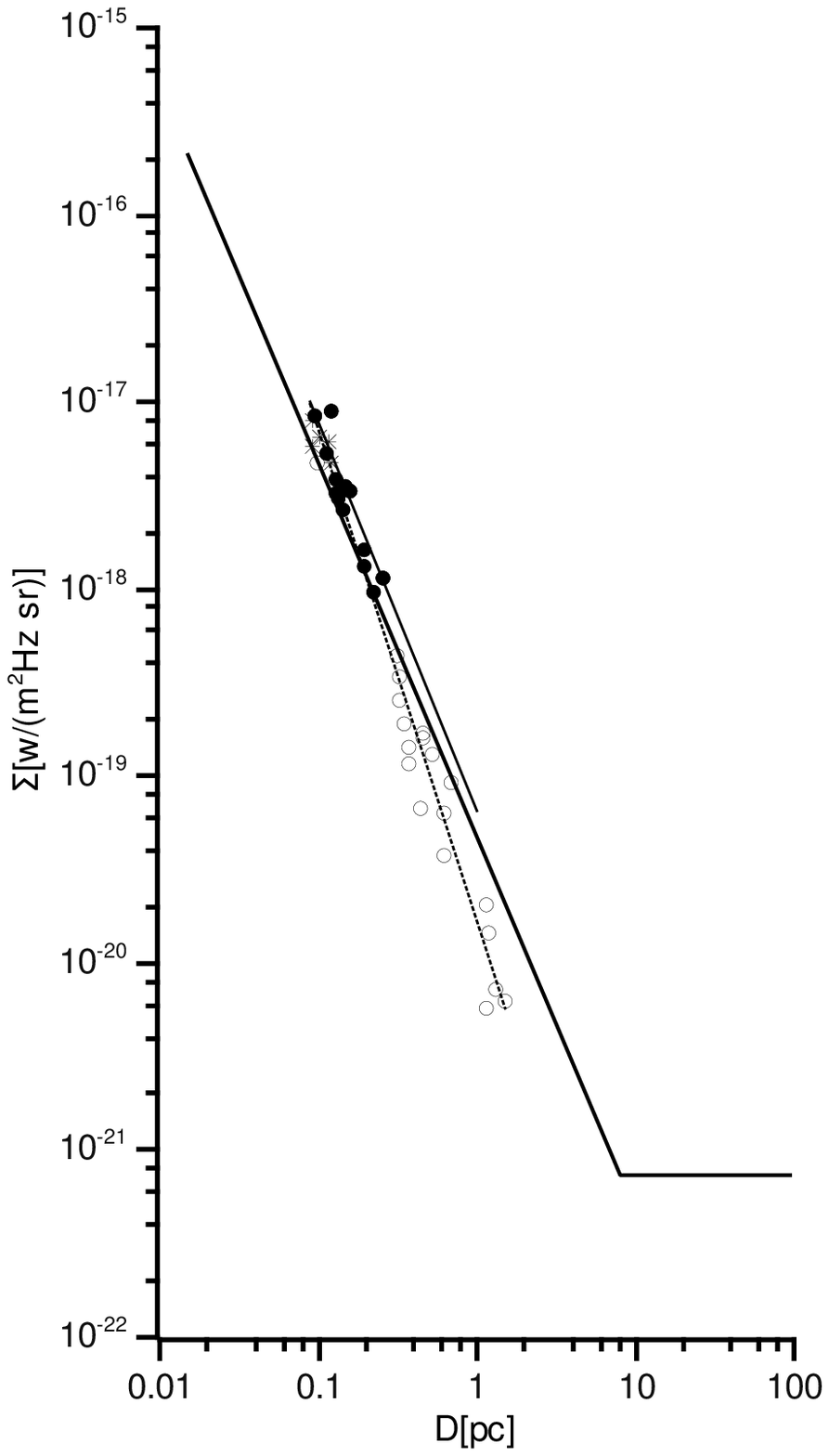}
\caption{The Monte Carlo generated sample at {8.64}~GHz for a scatter of {2} and simulated slope of {--2.7}. The LMC data points (asterisks) are plotted along with the sensitivity ({thick} solid) line; artificially generated points are plotted above (filled circles) and below (open circles) this line. The selection of points above the sensitivity line give a fit with a flatter slope ({thin solid} line) than the slope of a fit using all simulated points (dotted line).}
\label{Fig2}
\end{figure*}

In Tables~\ref{48ghz} and~\ref{864ghz}, the first column lists the value of the simulated slope, while the mean and standard deviation of the best fit slopes for the generated samples are given in second and third column, respectively. The fourth and fifth column gives the mean and standard deviation of the best fit slopes for sensitivity selected generated samples, respectively. In Fig.~\ref{Fig2} we present one of our Monte Carlo generated samples at {8.64}~GHz for a scatter of 1 and the simulated slope {--2.7}.

\begin{table}
\tiny
\caption{Monte Carlo simulation results at 4.8 GHz.} \label{48ghz}
\begin{tabular}{@{}ccccc@{\hspace{5pt}}c@{}}
\hline

The slope&Mean&Standard&Mean&Standard&No. of\\
that is&simulated&deviation&slope&deviation&generated\\
simulated&slope&of mean&after&of slope&samples\\
&&simulated&selection&after&that was\\
&&slope&&selection&fitted\\
\multicolumn{6}{c}{}\\
\hline
\multicolumn{6}{l}{scatter=1.0}\\
--1.500000& --1.492251& 0.068066& --1.514832& 0.064286& 100\\
--1.600000& --1.598043& 0.068590& --1.623227& 0.062590& 100\\
--1.700000& --1.699330& 0.065987& --1.728391& 0.064682& 100\\
--1.800000& --1.801370& 0.061776& --1.827670& 0.062875& 100\\
--1.900000& --1.904920& 0.065998& --1.939695& 0.063711& 100\\
--2.000000& --2.003822& 0.057447& --2.001172& 0.059317& 100\\
--2.100000& --2.099951& 0.065622& --2.041389& 0.062283& 100\\
--2.200000& --2.195110& 0.063386& --2.070956& 0.086876& 99\\
--2.300000& --2.297590& 0.064135& --2.074784& 0.130676& 82\\
--2.400000& --2.398137& 0.070116& --2.110495& 0.211172& 59\\
--2.500000& --2.501291& 0.065666& --2.180770& 0.426777& 40\\
--2.600000& --2.594927& 0.068783& --2.222386& 0.527765& 23\\
--2.700000& --2.700259& 0.065226& --2.347649& 0.527067& 11\\
--2.800000& --2.801124& 0.071673& --2.339542& 0.345267& 11\\
--2.900000& --2.892171& 0.061980& --2.325770& 0.384151& 9\\
--3.000000& --2.995751& 0.067793& --2.814787& 0.608717& 4\\
--3.100000& --3.093523& 0.066956& --2.789218& 0.225177& 3\\
--3.200000& --3.093523& 0.131417& -- & -- &0\\
--3.300000& --3.304876& 0.071690& --2.841456& 0.077769& 3\\
--3.400000& --3.304876& 0.126657& --2.841456& -- & 1\\
--3.500000& --3.502017& 0.064626& --2.471982& 0.471662& 3\\

\multicolumn{6}{c}{}\\
\hline
\multicolumn{6}{l}{scatter=2.0}\\

--1.500000& --1.494408& 0.129362& --1.582530& 0.129317& 100\\
--1.600000& --1.582302& 0.137065& --1.684048& 0.127792& 100\\
--1.700000& --1.710318& 0.142266& --1.786533& 0.126412& 100\\
--1.800000& --1.803957& 0.135674& --1.896290& 0.127258& 100\\
--1.900000& --1.908844& 0.128375& --1.957930& 0.123818& 100\\
--2.000000& --1.977503& 0.128244& --1.979623& 0.101482& 100\\
--2.100000& --2.126453& 0.124008& --2.047942& 0.106982& 100\\
--2.200000& --2.218740& 0.132994& --2.069085& 0.121897& 99\\
--2.300000& --2.301067& 0.118757& --2.100132& 0.180063& 96\\
--2.400000& --2.422463& 0.156185& --2.121199& 0.220895& 85\\
--2.500000& --2.489904& 0.122634& --2.137602& 0.234918& 73\\
--2.600000& --2.596131& 0.130540& --2.133221& 0.314863& 54\\
--2.700000& --2.715585& 0.105664& --2.166716& 0.624909& 38\\
--2.800000& --2.784944& 0.118933& --2.127692& 0.398607& 30\\
--2.900000& --2.922056& 0.123773& --2.334104& 0.607681& 16\\
--3.000000& --3.007808& 0.115488& --2.198724& 1.185409& 20\\
--3.100000& --3.090180& 0.137786& --2.386109& 0.311125& 10\\
--3.200000& --3.181103& 0.140162& --2.454902& 1.599802& 6\\
--3.300000& --3.319871& 0.139671& --1.996534& 0.496468& 9\\
--3.400000& --3.389912& 0.124211& --2.349205& 0.254864& 4\\
--3.500000& --3.389912& 0.168488& --2.349205& -- & 1\\

\hline
\end{tabular}

 \end{table}

\begin{table}
\tiny
\caption{Monte Carlo simulation results at 8.64 GHz.} \label{864ghz}
\begin{tabular}{@{}ccccc@{\hspace{5pt}}c@{}}

\hline

The slope&Mean&Standard&Mean&Standard&No. of\\
that is&simulated&deviation&slope&deviation&generated\\
simulated&slope&of mean&after&of slope&samples\\
&&simulated&selection&after&that was\\
&&slope&&selection&fitted\\
\multicolumn{6}{c}{}\\
\hline
\multicolumn{6}{l}{scatter=1.0}\\
--1.500000& --1.503405& 0.052604& --1.503405& 0.052604& 100\\
--1.600000& --1.600591& 0.042812& --1.601002& 0.042035& 100\\
--1.700000& --1.698601& 0.037607& --1.698601& 0.037607& 100\\
--1.800000& --1.792767& 0.040580& --1.792767& 0.040580& 100\\
--1.900000& --1.893138& 0.043315& --1.893138& 0.043315& 100\\
--2.000000& --1.995468& 0.045243& --1.994920& 0.045646& 100\\
--2.100000& --2.102842& 0.042350& --2.088784& 0.045885& 100\\
--2.200000& --2.193731& 0.042313& --2.144698& 0.044331& 100\\
--2.300000& --2.298443& 0.045962& --2.187980& 0.069080& 100\\
--2.400000& --2.401427& 0.048780& --2.269207& 0.107564& 100\\
--2.500000& --2.501493& 0.043743& --2.318480& 0.162147& 98\\
--2.600000& --2.601538& 0.043349& --2.425642& 0.195200& 98\\
--2.700000& --2.699132& 0.040760& --2.449996& 0.249949& 81\\
--2.800000& --2.803434& 0.040787& --2.553920& 0.368122& 62\\
--2.900000& --2.895824& 0.040305& --2.556149& 0.549145& 56\\
--3.000000& --3.000702& 0.041717& --2.694128& 0.306140& 49\\
--3.100000& --3.100702& 0.039240& --2.703438& 0.557734& 40\\
--3.200000& --3.199087& 0.044836& --2.836055& 0.529287& 28\\
--3.300000& --3.303262& 0.042204& --2.874302& 0.325416& 15\\
--3.400000& --3.394428& 0.044995& --2.775834& 0.499802& 18\\
--3.500000& --3.495130& 0.043181& --2.928748& 0.433336& 14\\

\multicolumn{6}{c}{}\\
\hline
\multicolumn{6}{l}{scatter=2.0}\\

--1.500000& --1.499001& 0.073477& --1.506813& 0.071878& 100\\
--1.600000& --1.595906& 0.091924& --1.609181& 0.091654& 100\\
--1.700000& --1.706313& 0.079195& --1.720427& 0.077824& 100\\
--1.800000& --1.801306& 0.077013& --1.818025& 0.079255& 100\\
--1.900000& --1.892254& 0.092443& --1.903992& 0.083989& 100\\
--2.000000& --2.002529& 0.093860& --2.007141& 0.092602& 100\\
--2.100000& --2.094921& 0.087675& --2.065955& 0.079765& 100\\
--2.200000& --2.196983& 0.097031& --2.109052& 0.089702& 100\\
--2.300000& --2.278865& 0.094011& --2.135513& 0.085328& 100\\
--2.400000& --2.403615& 0.089715& --2.205754& 0.141621& 100\\
--2.500000& --2.478826& 0.080478& --2.194225& 0.191552& 98\\
--2.600000& --2.593142& 0.095069& --2.264253& 0.276645& 89\\
--2.700000& --2.709181& 0.091667& --2.402365& 0.329761& 85\\
--2.800000& --2.799243& 0.076143& --2.297942& 0.374669& 67\\
--2.900000& --2.901775& 0.080564& --2.406221& 0.411391& 61\\
--3.000000& --2.999507& 0.086134& --2.591900& 0.528155& 41\\
--3.100000& --3.083982& 0.090140& --2.527112& 0.497492& 31\\
--3.200000& --3.187697& 0.081336& --2.453453& 0.617842& 29\\
--3.300000& --3.303730& 0.089221& --2.681570& 1.138125& 32\\
--3.400000& --3.392252& 0.084091& --2.712006& 0.649541& 20\\
--3.500000& --3.501420& 0.078870& --2.584415& 0.885438& 21\\

\hline
\end{tabular}

 \end{table}

%___________________________________________________________________

\section{Discussion and conclusions}

From tables \ref{48ghz} and \ref{864ghz}, it is evident that the sensitivity cutoff tends to flatten the intrinsic (simulated) slopes less than --2 (the most likely case for the real data). {At 4.8~GHz, a slope of $-0.9\pm0.5$ (shallower than the sensitivity line) is not influenced with sensitivity related selection effects. From the other side, this slope does not have any physical interpretation and is only the consequence of scatter in the $\Sigma-D$ plane. The 8.64 GHz slope of $-2.6\pm0.4$ appears to correspond to somewhat steeper selection free slopes but due to a large error could have any value smaller than  --2.3. This is within the range of the SSV slope at 4.8 GHz. With current LMC data samples it is not possible to constrain the lower limit due to a smaller number of selected points for steeper slopes. The value of a mean slope after selection at 8.64 GHz starts to oscillate for the slopes $\lesssim -2.4$ (the standard deviation of slope after selection exceeds the interval between two successive simulated slopes). Table \ref{48ghz} is given here rather for completeness, but nevertheless, shows that for the samples with small number of data points and a slope larger than --2 it is not possible to extract any meaningful information with this kind of simulations.} 

With these {recently confirmed MCs radio} PNe being amongst brightest Galactic PNe in the $\Sigma-D$ plane and given their proximity to the sensitivity line, it is likely they represent only the peak of the actual {MCs} PNe luminosity function, {which is probably the ''bright-end'' extension of the Galactic PNe luminosity function \citep[for a more detailed discussion on the nature of these MCs PNe see][]{filipovic_etal09}}. {The results of our LMC Monte Carlo simulations suggest that current sparse data samples cannot give meaningful $\Sigma-D$ slope. However, they confirmed that the corresponding apparent $\Sigma-D$ slope flattens toward $\beta \approx -2$ because of the sensitivity related selection effects.}. 

{This should be kept in mind as a serious measure of caution in forthcoming studies of extragalactic PN samples.} Future high sensitivity images of the MCs will surely provide even better and more complete radio samples of PN population {and enable more robust constraining of PN evolution parameters.}

\begin{acknowledgements}
{The authors would like to thank the referee Prof. John Dickel for useful comments that have improved the manuscript.} This research has been supported by the Ministry of Science and Technological Development of the Republic of Serbia through the Project \#146012.

\end{acknowledgements}

\bibliographystyle{aa}
        \bibliography{mybib}

\end{document}